\documentstyle[sprocl,psfig]{article}

\def\Journal#1#2#3#4{{#1} {\bf #2}, #3 (#4)}

\def\HIP{\em Heavy Ion Phys.}
\def\NPB{{\em Nucl. Phys.} B}
\def\NPA{{\em Nucl. Phys.} A}
\def\PLB{{\em Phys. Lett.}  B}
\def\PRL{\em Phys. Rev. Lett.}

\def\PRC{{\em Phys. Rev.} C}
\def\ZPC{{\em Z. Phys.} C}

\newcommand{\insertplotshot}[1]{
\centerline{\psfig{figure={#1},height=9.0cm}}
}

\newcommand{\beq}{\begin{equation}}
\newcommand{\eeq}[1]{\label{#1} \end{equation}}
\newcommand{\beqar}{\begin{eqnarray}}
\newcommand{\eeqar}[1]{\label{#1} \end{eqnarray}}
\newcommand{\dlt}{\bigtriangleup}

\begin{document}

\title{The source of the "third flow component"}

\author{V. K. Magas\footnote{Talk given at the 30th International 
Symposium on Multiparticle Dynamics, October 9-15, 2000, Tihany, 
Lake Balaton, Hungary
}, L. P. Csernai }

\address{Section for Theoretical and Computational Physics, 
Department of Physics\\University of
Bergen, Allegaten 55, 5007 Bergen, Norway\\
E-mail: vladimir@fi.uib.no, csernai@fi.uib.no} 

\author{D. D. Strottman}

\address{Theoretical Division, Los Alamos National Laboratory\\
 Los Alamos, NM, 87454, USA\\
 E-mail: dds@lanl.gov}


\maketitle\abstracts{A model for energy, pressure and flow velocity 
distributions at the beginning of relativistic heavy ion
collisions has been developed to be used as initial condition for 
hydrodynamical calculations. The results
show that at the early stages QGP forms a tilted disk, such that 
the direction of the largest pressure gradient
stays in the reaction plane, but deviates from both the beam and 
the usual transverse flow directions. Such
initial conditions may lead to the creation of the "antiflow" or 
the "third flow component" \cite{CR}.}

\section{Introduction}\label{int}
Fluid dynamics is probably the most frequently used model to
describe heavy ion collisions. It assumes local equilibrium, i.e. the
existence of an Equation of State (EoS), relatively short range
interactions and conservation of energy and momentum as well as
of conserved charge(s).
Thus, in ultra relativistic collisions of 
large heavy ions, especially if Quark-Gluon Plasma (QGP) is formed, one-fluid 
dynamics is a valid and good description for the intermediate stages of the reaction. Here 
interactions are strong and frequent, so that other models, (e.g. transport models, string 
models, etc., assuming binary collisions, with free propagation of constituents between collisions) 
have limited validity. On the other hand, the initial and final, Freeze-Out (FO), stages of the reaction
are outside the domain of applicability of the fluid dynamical model.

We believe that the realistic and detailed description of an energetic heavy ion reaction 
requires a
 Multi Module Model, where the different stages of the reaction are each described with suitable 
theoretical approach. It is important that these Modules are coupled to each other correctly: 
on the interface, which is a 3 dimensional hyper-surface in space-time, 
all conservation laws should be satisfied and entropy should 
not decrease. These matching conditions were worked out and studied 
for the matching at FO in detail in refs.  \cite{FO2,FO3}.

The final FO stages of the reaction, after hadronization, can be described well with kinetic models,
since the matter is already dilute.

The initial stages are more problematic. Frequently two or three fluid models are used to remedy 
the difficulties and to model the process of QGP formation and thermalization \cite{A78,C82,bsd00}.
Here the problem is transferred to the determination of drag-, friction- and transfer- terms among the 
fluid components, and a new problem is introduced with the (unjustified) use of EoS in each component 
in nonequilibrated situations, where EoS does not exist. Strictly speaking this approach can only be 
justified for mixtures of noninteracting ideal gas components.  Similarly, the use of transport theoretical
approaches assuming dilute gases with binary interactions is questionable, since, due to the extreme 
Lorentz contraction in the C.M. frame, enormous particle and energy densities with the immediate 
formation of perturbative vacuum should be handled. Even in most parton cascade models these initial 
stages of the dynamics are just assumed in form of some initial condition
with little justification behind.

Our goal in the present work is to construct a model, based on the recent 
experience gained in string Monte Carlo models and in parton cascades, describing
energy, pressure and flow velocity distributions at the beginning of relativistic heavy ion
collisions.  

\section{The effective string rope model}\label{model}
One of the important results of the last decade is that 
all string models had to introduce new, energetic objects:
string ropes \cite{bnk84,S95}, quark clusters \cite{WA96}, fused strings \cite{ABP93},
in order to describe the abundant formation of massive particles like strange antibaryons. 
Based on this, we describe the initial stages of the reaction in the framework of classical 
(or coherent) Yang-Mills theory, generalizing ref. \cite{GC86} assuming larger field strength 
(string tension) than in ordinary hadron-hadron collisions.  In addition we now satisfy all 
conservation laws exactly, while in ref.  \cite{GC86} infinite projectile energy was assumed, 
and so, overall energy and momentum conservation was irrelevant. 

First of all, we would create a grid in $[x,y]$ plane
($z$ -- is the beam axes, $[z,x]$ -- is the reaction plane) and 
describe the nucleus-nucleus collision in terms of steak-by-streak collisions, 
corresponding to the same transverse coordinates, $\{x_i, y_j\}$.  We assume that baryon 
recoil for both target and projectile arise from the acceleration of partons in an effective field 
$F^{\mu\nu}$, produced in the interaction.  Of course, the physical picture behind this model 
should be based on chromoelectric flux tube or string models, but for our purpose we consider
 $F^{\mu\nu}$ as an effective abelian field. A single phenomenological parameter describing this field 
must be fixed from comparison with experimental data.
For detailed description we would call 
the attention of the reader to refs. \cite{MCS_rev,MCS00,CAM00}. 

Briefly speaking, after two streaks cross each other they create a "string rope" 
with constant string tension 
$\sigma$ in the middle. The typical values of the string
tension, $\sigma$, are of the order of $10\ GeV/fm$, and these may be
treated as several parallel strings or "string rope". 
In further evolution this "string rope" accommodates more and more energy 
by stopping the initial target and projectile streaks. If we let system evolve, 
the "string rope" will stop streaks completely, 
then will accelerate them backward, and the analog of Yo-Yo motion may occur.    
We do not solve 
simultaneously the kinetic problem leading to parton equilibration, just assume that the 
arising friction is such that the heavy ion system will be an overdamped oscillator, i.e.
yo-yoing of the two heavy ions will not occur. This assumption is based on recent string 
and parton cascade results. 

To proceed one has to make more assumptions. 
We assume that the final result of collisions of two streaks, after
stopping the string's expansion and after its decay, is one streak of the
length $\dlt l_f$ with homogeneous energy density distribution, $e_f$, and
baryon charge distribution, $\rho_f$, moving like one object with rapidity
$y_f$. We assume that this is due to string-string interactions and string
decays. The homogeneous distributions are the simplest assumptions, which may be
modified based on experimental data. Its advantage is a simple expression
for $e_f,\ \rho_f,\ y_f$.

The final energy density, baryon density and rapidity, $e_f, \ \rho_f$ and
$y_f$, should be determined from conservation laws. Unfortunately, the 
assumptions we made above oversimplify the real situations and do not 
allow us to satisfy exactly all the conservation laws \cite{MCS_rev,CAM00}. 
The reason for this is well known and has been discussed in the 
Refs. \cite{FO2,FO3}:  
two possible definitions of the flow, Eckart's and Landau's definition.
If we are following the energy
flow, we satisfy exactly the energy and momentum conservation  and 
violate the baryon current conservation. Otherwise if we are drifted 
by baryon flow, we
violate the energy-momentum conservation. 

In the refs.  \cite{MCS_rev,MCS00,CAM00} we choose the Landau's definition of the $y_f$, 
i.e. the exact conservation of the energy and momentum.

\begin{figure}[htb]
\insertplotshot{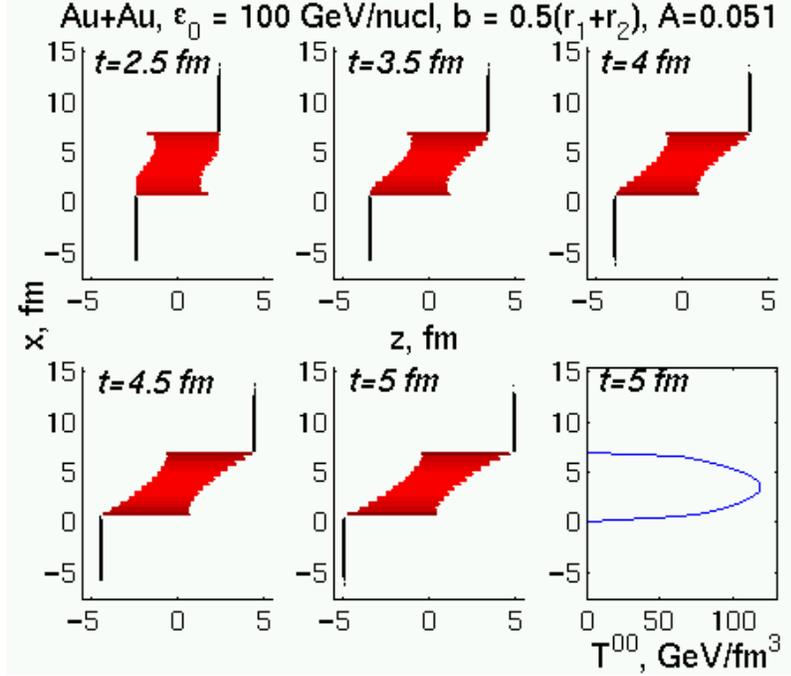}
\caption[]{The Au+Au collisions, $\varepsilon_0=100\ GeV/nucl$, 
$b=0.5(R_1+R_2)$, $A=0.051$ (parameter $A$ introduced in refs. \cite{MCS_rev,MCS00,CAM00} 
to determine the "string rope" tension),
 $y=0$ (ZX plane through 
the centers of nuclei). We would like to notice that 
final shape of QGP volume is a tilted disk $\approx 45^0$, 
and the direction of the fastest expansion will deviate from both 
the beam axis and the usual transverse flow direction, and might be a 
reason for the third flow component, as argued in \cite{CR}.
\label{ev11}}
\end{figure}

\section{Initial state of QGP}\label{initial}
Let us review the results of our calculations. We are interested in the shape of 
QGP formed, when the final streak formation is finished (at least in the most 
central region \cite{MCS_rev}) and their matter is 
locally equilibrated. 
This will be the initial state for further hydrodynamical calculations. 
The time, where we expect a local equilibrium, is also a parameter of our model - 
it should be large enough to allow the final streak formation, but we can't wait too long, 
since the transverse expansion can't be neglected in this case.

\begin{figure}[htb]
\insertplotshot{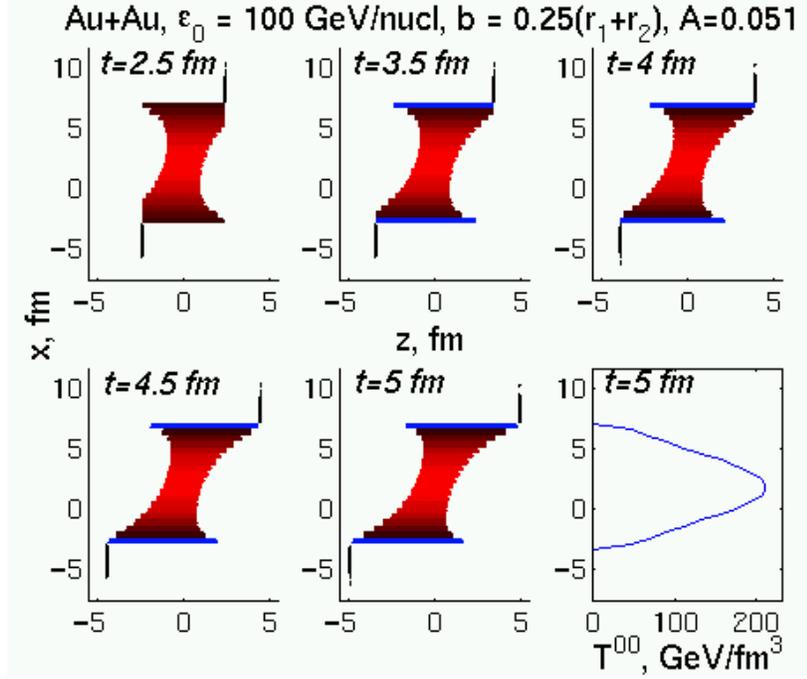}
\caption[]{The same as Fig. \ref{ev11}, but $b=0.25(r_1+r_2)$. We see that 
for more central collisions the energy density is much larger. 
The QGP volume 
has a shape of tilted disk and may produce a third flow component 
\cite{CR}. }

\label{ev12}
\end{figure}

We see in 
Figs. \ref{ev11}, \ref{ev12} that QGP forms a tilted disk for $b\not =0$. So, 
the direction of fastest 
expansion, the same as largest pressure gradient, 
will deviate from both the beam axis
and the usual transverse flow direction and the 
new flow component, called "antiflow" or "third flow component", 
may appear in the reaction plane. 
With increasing beam energy the usual transverse flow is getting 
weaker, while this new flow component 
is strengthened. The mutual effect of the usual directed transverse 
flow and this new "antiflow" or 
"third flow component" leads to an enhanced emission in 
the reaction plane.  This was actually 
observed and widely studied earlier 
and was referred to as "elliptic flow".

\section{Conclusions}\label{concl}
Based on earlier Coherent Yang-Mills field theoretical models and introducing effective 
parameters based on Monte-Carlo string cascade and parton cascade model results, a 
simple model is introduced to describe the pre fluid dynamical stages of heavy ion 
collisions at the highest SPS energies and above.  

Contrary to earlier expectations, --- based on standard string tensions of $1\ GeV/fm$ 
which lead to the Bjorken model type of initial state, --- the effective string tension 
introduced for collisions of massive heavy ions, as a consequence of collective 
effects related to QGP formation, appears to be of the order of $10\ GeV/fm$ and consequently 
causes much less transparency. The resulting initial locally 
equilibrated state of matter in semi central collisions takes a rather unusual form, which 
can be then identified by the asymmetry of the caused collective flow. 

Detailed fluid dynamical calculations as well as flow experiments at
semi central impact parameters for massive heavy ions are needed 
at SPS and RHIC energies to connect the predicted special initial state
with observables.



\end{document}